\documentclass[12pt]{article}
\usepackage{graphicx}
\usepackage{siunitx}
\usepackage[italic]{hepnames}
\usepackage{amsmath}
\usepackage{amsthm}
\usepackage{xspace}
\usepackage{enumitem}
\usepackage{lineno}

\graphicspath{{./}}

\newcommand\pubnumber{SNSN-323-63}
\newcommand\pubdate{\today}

\def\institute{Physikalisches Institut, University of Bonn, Bonn, Germany}
\def\support{\footnote{Copyright [2018] CERN for the benefit of the [ATLAS Collaboration]. CC-BY-4.0 license.}}

\def\Title#1{\begin{center} {\Large #1 }
\end{center}}
\def\Author#1{\begin{center}{ \sc #1}
\end{center}}
\def\Address#1{\begin{center}{ \it #1}
\end{center}}

\newcommand\pubblock{\rightline{\begin{tabular}{l}
\pubnumber\\
         \pubdate  \end{tabular}}}
\newenvironment{Abstract}{\begin{quotation} }{\end{quotation}}
\newenvironment{Presented}{\begin{quotation}
\begin{center} PRESENTED AT\end{center}\bigskip
      \begin{center}\begin{large}}{\end{large}\end{center}
      \end{quotation}}

\newcommand*{\TeV}{\ensuremath{\text{Te\kern -0.1em V}}}
\newcommand*{\GeV}{\ensuremath{\text{Ge\kern -0.1em V}}}
\newcommand*{\ttbar}{\Ptop{}\APtop}
\newcommand*{\pt}{\ensuremath{p_{\text{T}}}\xspace}
\newcommand*{\lumiValue}{36.1}
\newcommand*{\ifb}{\mbox{fb$^{-1}$}}
\newcommand*{\lumiUnit}{\ifb}

\newcommand*{\lumi}{\SI[parse-numbers=false]{\lumiValue}{\lumiUnit}\xspace}
\newcommand*{\met}{\ensuremath{E_{\text{T}}^{\text{miss}}}\xspace}
\newcommand*{\Zboson}{\ensuremath{Z}\xspace}
\newcommand*{\Zjets}{\ensuremath{\Zboson\text{\,+\,jets}}\xspace}
\newcommand*{\EJet}{\ensuremath{E(b)}\xspace}
\newcommand*{\ELepLepJet}{\ensuremath{E(\ell \ell b)}\xspace}
\newcommand*{\massTransLepLepMETJet}{\ensuremath{m_{\textrm{T}} (\ell\ell\nu\nu b)}\xspace}
\newcommand*{\massLepOneJet}{\ensuremath{m(\ell_1 b)}\xspace}
\newcommand*{\massLepLepJet}{\ensuremath{m(\ell \ell b)}\xspace}
\newcommand*{\massLepTwoJet}{\ensuremath{m(\ell_2 b)}\xspace}
\newcommand*{\POWHEGBOX}{\textsc{Powheg-Box}\xspace}
\newcommand*{\HERWIGpp}{Herwig++\xspace}
\newcommand*{\PYTHIAV}[1]{\textsc{Pythia}~#1\xspace}

\begin{document}
\begin{titlepage}
\pubblock

\vfill
\Title{Measurement of $tW$ differential cross-sections with ATLAS at $\sqrt{s}=\SI{13}{\TeV}$}
\vfill
\Author{ Rui Zhang, on behalf of the ATLAS Collaboration\support}
\Address{\institute}
\vfill
\begin{Abstract}
      The cross-section to produce a $W$ boson in association with a top quark is measured
      differentially with respect to several particle-level final-state observable quantities. The
      measurements are performed using $36.1$~\mbox{fb$^{-1}$} of $pp$ collision data at
      $\sqrt{s}=13~\mathrm{TeV}$ collected in 2015 and 2016, by the ATLAS detector at the LHC. Cross-sections are measured in a
      fiducial phase-space defined by the presence of two charged leptons and exactly one jet
      identified as containing $B$ hadrons. Measurements are normalised to the fiducial
      cross-section, causing several of the main uncertainties to cancel. The results are found to be in
      good agreement with predictions from several Monte Carlo generators.
\end{Abstract}
\vfill
\begin{Presented}
$10^{th}$ International Workshop on Top Quark Physics\\
Braga, Portugal,  September 17--22, 2017
\end{Presented}
\vfill
\end{titlepage}
\def\thefootnote{\fnsymbol{footnote}}
\setcounter{footnote}{0}

\section{Introduction}

Single-top-quark production via electroweak interactions involving a $Wtb$ vertex at leading order
is an excellent probe of the $Wtb$ couplings. Among all the possible mechanisms, the top-quark production in
association with a $W$ boson ($tW$) is the second largest process at the LHC. This cross-section has
been measured by the ATLAS~\cite{TOPQ-2015-16} and CMS~\cite{CMS-PAS-TOP-17-018} collaborations
using $\SI{13}{\TeV}$ collision data. These proceedings describe differential cross-section
measurements by the ATLAS collaboration in the $tW$ dilepton final state, which explore different kinematic regimes in a more
detailed way, and thus will be able to improve Monte Carlo (MC) modelling.

\section{Analysis strategy}

The data correspond to an integrated luminosity of \lumi{} at $\sqrt{s}=\SI{13}{\TeV}$ collected by ATLAS~\cite{PERF-2007-01}
in 2015 and 2016. Events are required to have exactly two oppositely
charged leptons (henceforth ``lepton'' refers to an electron or muon) with $\pt>\SI{27}{\GeV}$ and
$\pt>\SI{20}{\GeV}$, respectively, at least one of which has to be triggered on. Additionally,
events are required to have exactly one jet with $\pt>\SI{25}{\GeV}$ which is $b$-tagged ($b$-jet).
Finally a certain amount of missing transverse momentum, \met, is required depending on the invariant mass of two leptons, to
further reduce background from \Zjets{} events.

\begin{figure}[!h]
      \centering
      \includegraphics[width=.50\textwidth]{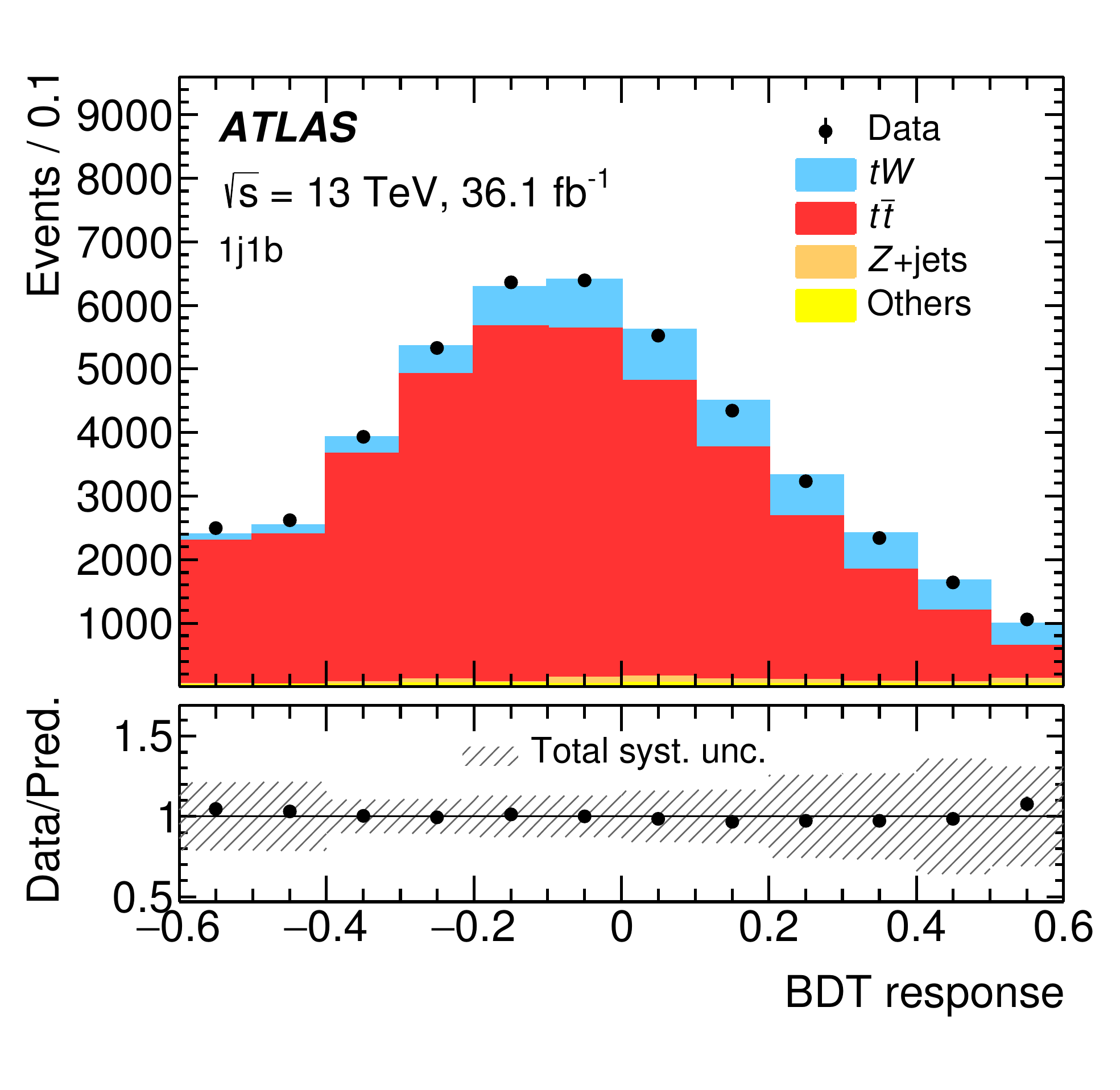}
      \caption{Comparison of data and MC predictions for the BDT response in the signal region. The
      $tW$ signal is normalised with the measured fiducial cross-section. Uncertainty bands reflect
      the total systematic uncertainties. The first and last bins contain underflow and overflow
      events, respectively \cite{TOPQ-2016-12}.}
      \label{fig:bdt}
\end{figure}

A boosted decision tree (BDT) technique~\cite{bdt} is used to combine several observables with increased
separation power into a single discriminant. The variables considered are derived from the kinematic
properties of subsets of the objects involved in the final states. The BDT discriminant distributions from MC predictions and data
are compared and shown in Figure~\ref{fig:bdt}.

To select a signal-enriched portion of events in the signal region, the BDT response is required to
be larger than $0.3$. The value of the requirement is optimised to reduce the total uncertainty of
the measurement over all bins, considering both statistical and systematic uncertainties.

The remained events are corrected for detector acceptance and resolution effects and the efficiency
to pass the event selection by using the iterative Bayesian unfolding
technique~\cite{DAgostini:1994zf} implemented in the \texttt{RooUnfold} software
package~\cite{Adye:2011gm}. The unfolding procedure includes bin-by-bin correction for
out-of-fiducial ($C^\text{oof}$) events which are reconstructed but fall outside the fiducial
acceptance at particle level, followed by the iterative matrix unfolding procedure $M^{-1}$, as well
as another bin-by-bin correction ($C^{\text{eff}}$) to the efficiency to reconstruct a fiducial
event:
\begin{equation*}
  N_i^{\text{ufd}} = \frac{1}{C_i^{\text{eff}}} \sum_j M_{ij}^{-1} C_j^{\text{oof}} (N_j^{\text{data}} - B_j),
\end{equation*}
where $N_i^\text{ufd}$ represents the unfolded event yields, $i$ ($j$) indicates the bin at particle (reconstruction) level, $N_j^{\text{data}}$ is the
number of events in data and $B_j$ is the sum of all background events.

Unfolded event yields are converted to cross-section values as a function of an observable $X$ using the expression:
\begin{equation*}
  \frac{\text{d}\sigma_i}{\text{d}X} =  \frac{N_i^{\text{ufd}}}{L \Delta_i},
\end{equation*}
where $L$ is the integrated luminosity of the data sample and $\Delta_i$ is the width of bin $i$ of the particle-level distribution.
Differential cross-sections are divided by the fiducial cross-section to create a normalised distribution.
The fiducial cross-section is simply the sum of the cross-sections in each bin multiplied by the corresponding bin widths:
\begin{equation*}
  \sigma^{\text{fid}} =  \sum_{i} \left( \frac{\text{d}\sigma_i}{\text{d}X} \cdot \Delta_i \right) = \sum_{i} \frac{N_i^{\text{ufd}}}{L}.
\end{equation*}

Many sources of experimental systematic uncertainties are taken into account. These include the
luminosity measurement, lepton efficiency scale factors used to correct simulation to data,
lepton/jet energy scale and resolution, \met related terms and the efficiency
of $b$-jet. The dominant systematic uncertainties in this category are related to the measurement of the jet
energy scale and resolution. Apart from the experimental systematics, uncertainties that arise due
to theoretical modelling of the signal and \ttbar{} background are also evaluated. The dominant
uncertainties for this analysis are the next-to-leading order (NLO) matrix element generator and the
parton shower and hadronisation generator.

\section{Results}

\begin{figure}[!htb]
      \begin{tabular}{ccc}
        \includegraphics[width=.31\textwidth]{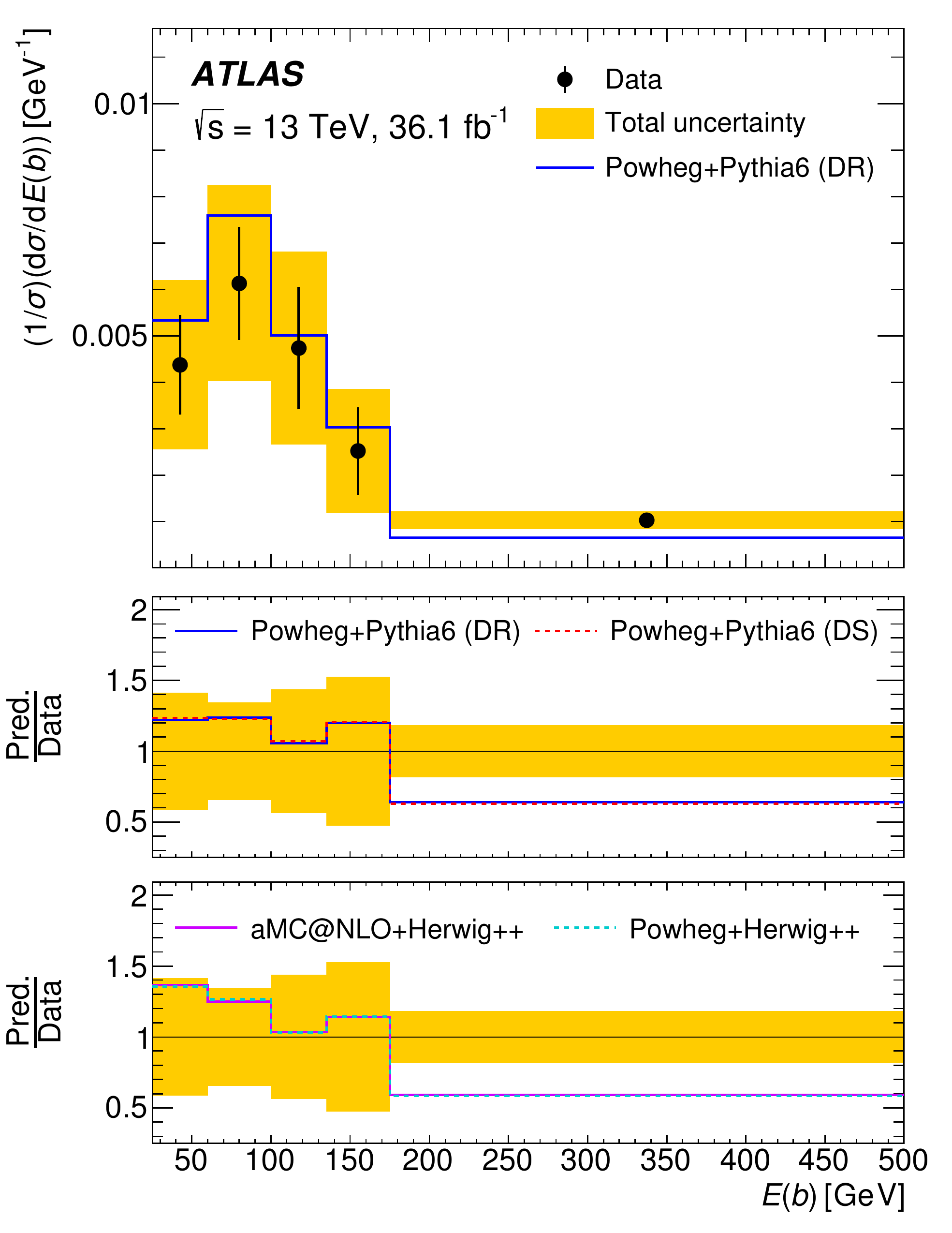} &
        \includegraphics[width=.31\textwidth]{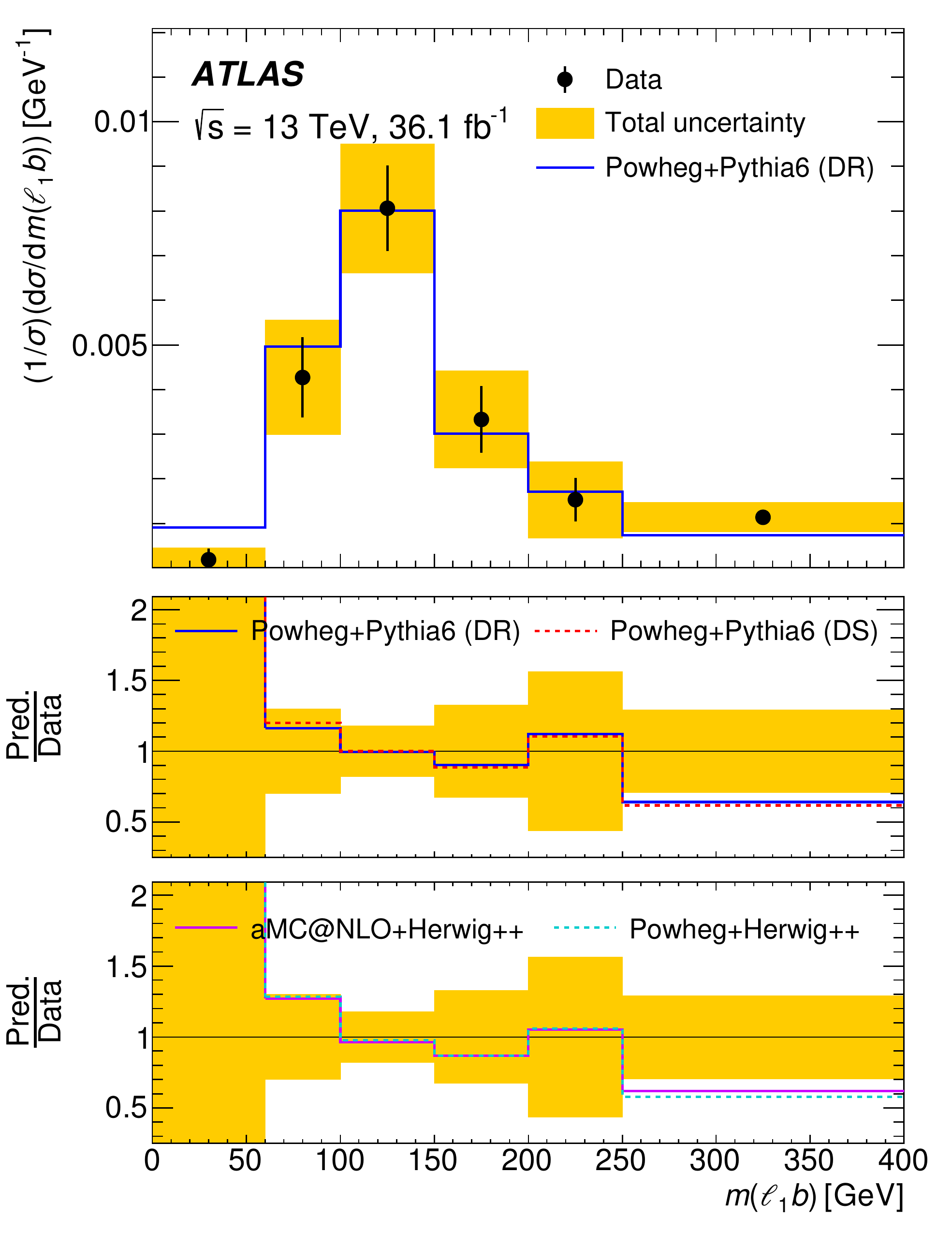} &
        \includegraphics[width=.31\textwidth]{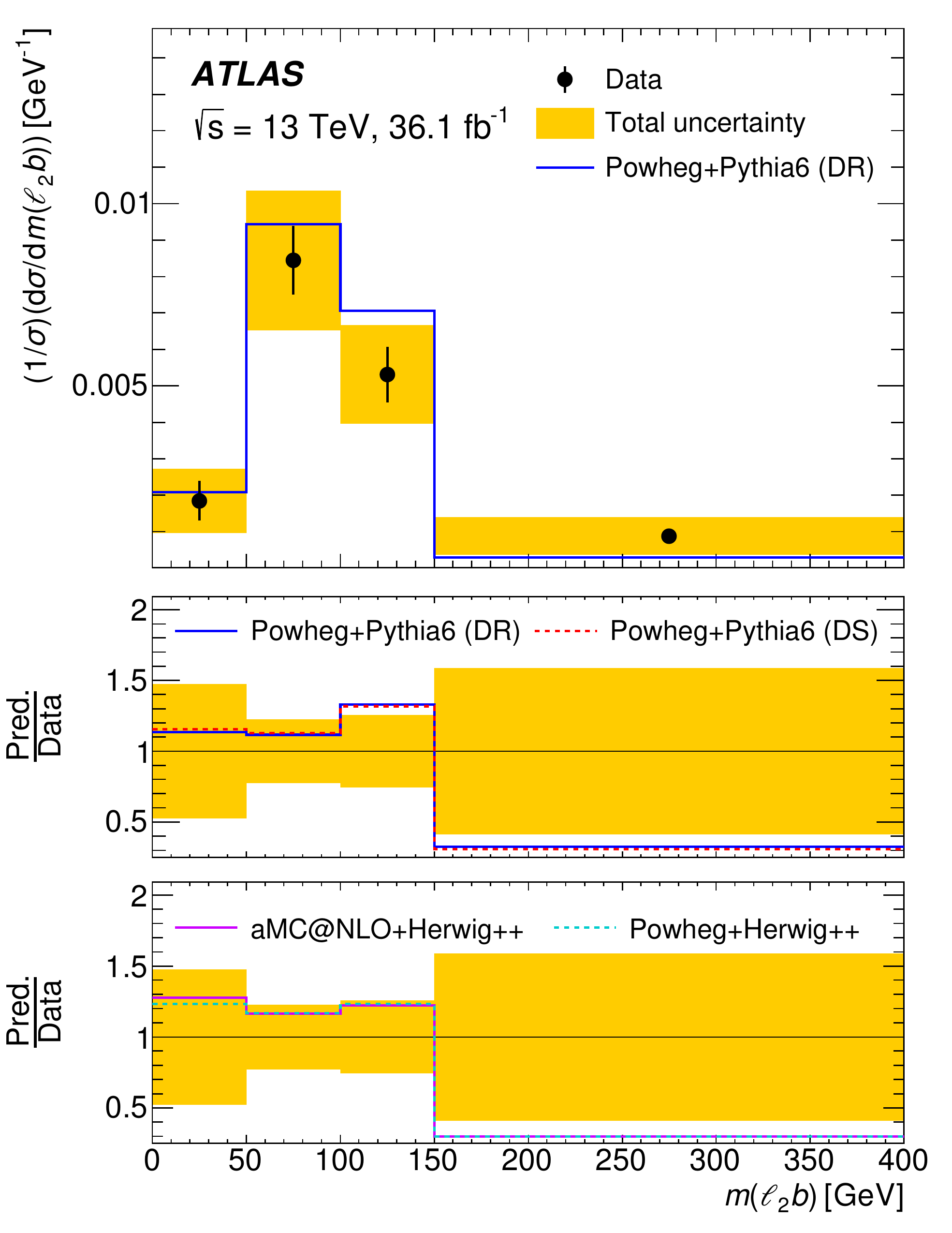} \\
        \includegraphics[width=.31\textwidth]{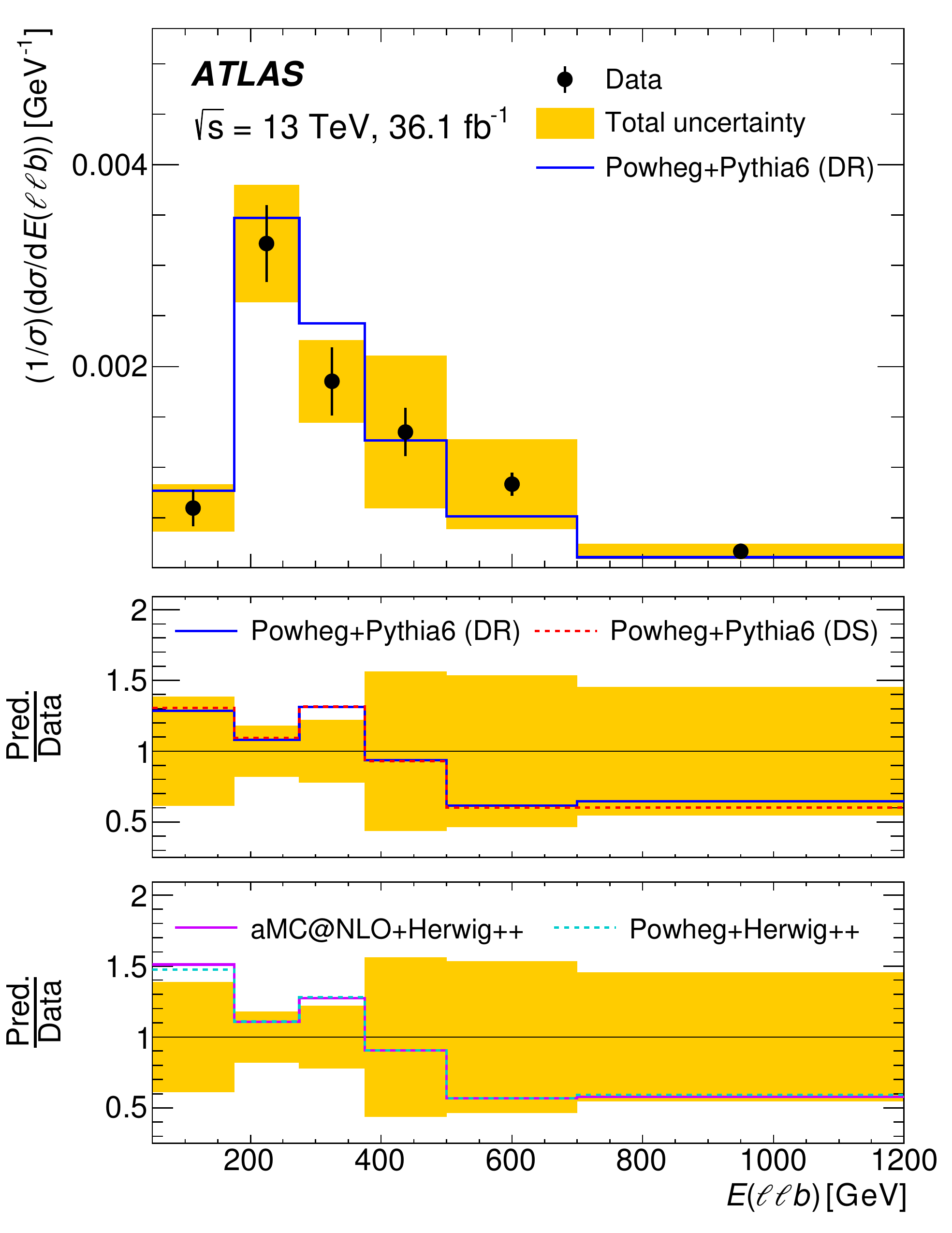} &
        \includegraphics[width=.31\textwidth]{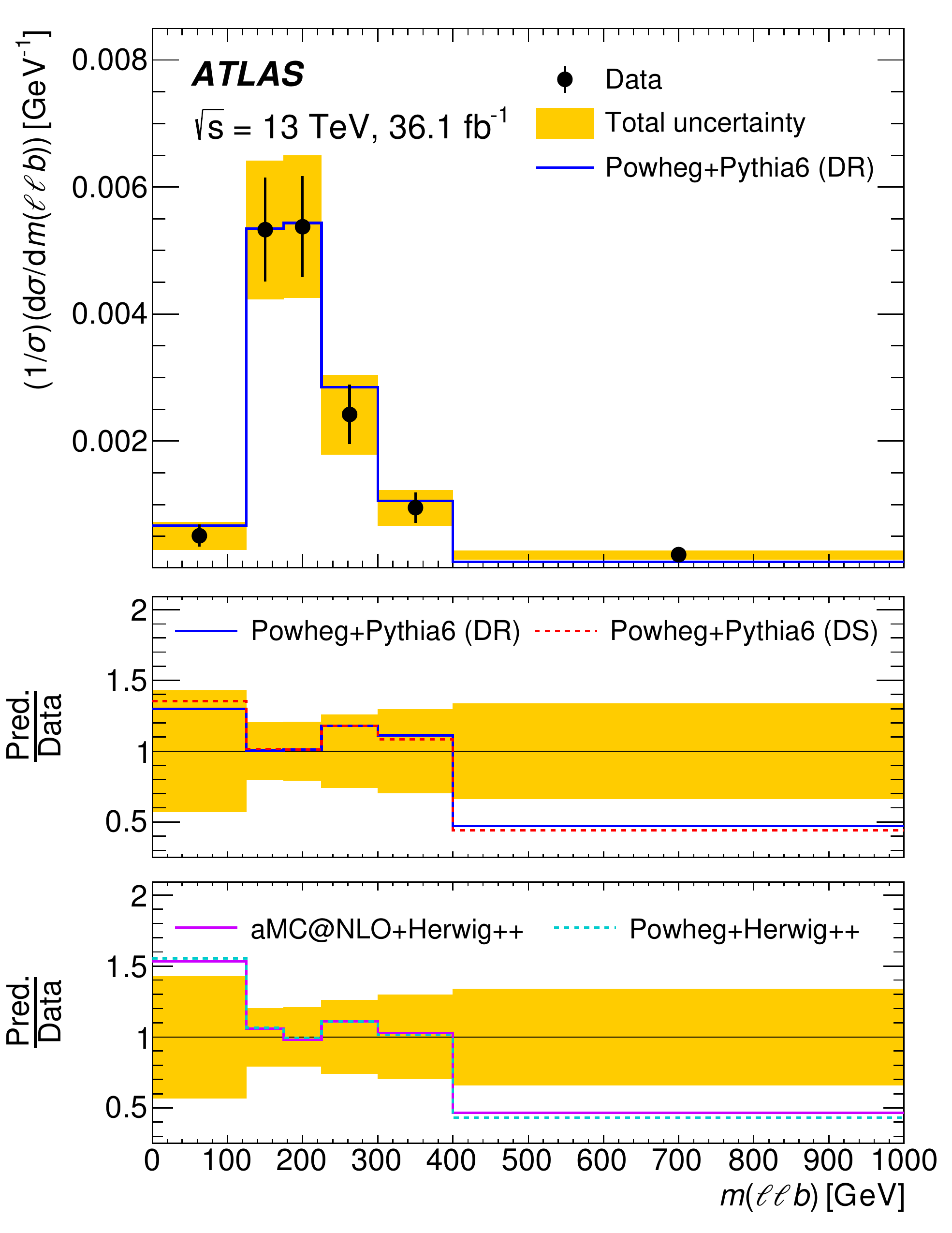} &
        \includegraphics[width=.31\textwidth]{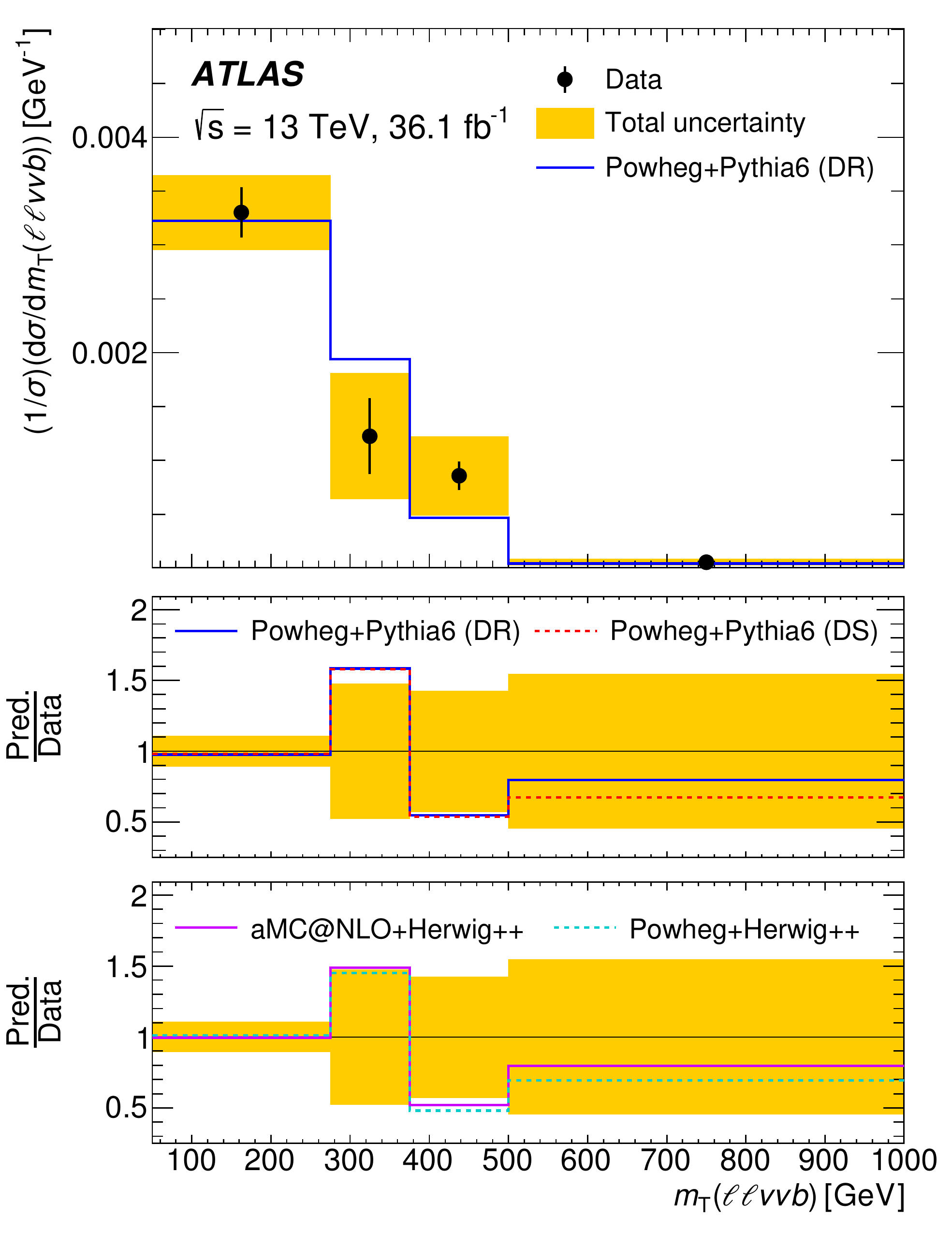} \\
      \end{tabular}
      \caption{\label{fig:result_norm1} Normalised differential cross-sections unfolded from data,
        compared with selected MC models, with respect to \EJet, \massLepOneJet, \massLepTwoJet, and
        \ELepLepJet. Data points are placed at the horizontal centre of each bin, and the error bars
        on the data points show the statistical uncertainties. The total uncertainty in the first
        bin of the \massLepOneJet{} distribution (not shown) is $140\%$ \cite{TOPQ-2016-12}.}
\end{figure}

Differential cross-sections are measured and compared to a variety of theory predictions (see Figure~\ref{fig:result_norm1}) for the
following variables:

\setlist{nolistsep}
\begin{itemize}[noitemsep]
      \item the energy of the $b$-jet, \EJet;
      \item the mass of the leading lepton and $b$-jet, \massLepOneJet;
      \item the mass of the sub-leading lepton and the $b$-jet, \massLepTwoJet;
      \item the energy of the system of the two leptons and $b$-jet, \ELepLepJet;
      \item the transverse mass of the leptons, $b$-jet and neutrinos, \massTransLepLepMETJet; and
      \item the mass of the two leptons and the $b$-jet, \massLepLepJet.
\end{itemize}
They are either related to the event, top quark or $W$ boson kinematics.

The uncertainty on the measurements is at the $20-50\%$ level. While this does not allow firm
conclusions to be drawn, in general, most of the MC models show fair agreement with the measured
cross-sections. Notably, for each distribution there is a substantial negative slope in the ratio of
predicted to observed cross-sections, indicating there are more events with high-momentum
final-state objects than several of the MC models predict. In most cases, differences between the MC
predictions are smaller than the uncertainty on the data, but there are some signs that
\POWHEGBOX{}+\HERWIGpp{} deviates more from the data and from the other predictions in certain bins
of the \ELepLepJet{}, \massLepLepJet{}, and
\massLepOneJet{} distributions. The predictions of DS and DR samples\footnote{Diagram removal (DR)
and diagram subtraction (DS) are two commonly used approaches to deal with quantum interference
between $tW$ and \ttbar{} processes.} likewise give very similar results for all observables as
expected from the fiducial selection. The predictions of \POWHEGBOX{}+\PYTHIAV{6} with varied
initial- and final-state radiation tuning were also examined but not found to give significantly
different distributions in the fiducial phase space of this analysis.


\begin{thebibliography}{99}


\bibitem{TOPQ-2015-16} ATLAS Collaboration, JHEP {\bf 01}, 063 (2018), arXiv:1612.07231 [hep-ex].

\bibitem{CMS-PAS-TOP-17-018} CMS Collaboration, CMS-PAS-TOP-17-018, (2017), arXiv:1710.03659 [hep-ex].

\bibitem{PERF-2007-01} ATLAS Collaboration, 2008 JINST {\bf 3}, S08003 (2008).

\bibitem{bdt} J. H. Friedman, Comput. Stat. \& Data Analysis {\bf 38}, 367 (2002).

\bibitem{TOPQ-2016-12} ATLAS Collaboration, Eur. Phys. J. C {\bf 78}, 186 (2018), arXiv:1712.01602 [hep-ex].

\bibitem{DAgostini:1994zf}
G. D'Agostini, Nucl. Instrum. Meth. A, {\bf 362}, 487 (1995).

\bibitem{Adye:2011gm}
T. Adye, arXiv: 1105.1160 [physics.data-an], (2011).

\end{thebibliography}
\end{document}